# Tool flank wear prediction using high-frequency machine data from industrial edge device

Deniz Bilgili [a] , Gamze Kecibas [a,b], Cemile Besirova [a,b], Mohammad Reza Chehrehzad [b], Gizem Burun [c], Toprak Pehlivan [a], Ugur Uresin [a], Engin Emekli [a], Ismail Lazoglu [b,*]

[a]*Ford Otosan R&D Center, Istanbul 34885, Turkey*
[b]*Koç University, Manufacturing and Automation Research Center, Istanbul 34450, Turkey*
[c]*Tubitak BILGEM Information Technologies Institute, Kocaeli 41400, Turkey*

* Corresponding author. Tel.: +90-212-3381587; *E-mail address:* ilazoglu@ku.edu.tr

**Abstract**

Tool flank wear monitoring can minimize machining downtime costs while increasing productivity and product quality. In some industrial applications, only a limited level of tool wear is allowed to attain necessary tolerances. It may become challenging to monitor a limited level of tool wear in the data collected from the machine due to the other components, such as the flexible vibrations of the machine, dominating the measurement signals. In this study, a tool wear monitoring technique to predict limited levels of tool wear from the spindle motor current and dynamometer measurements is presented. High-frequency spindle motor current data is collected with an industrial edge device while the cutting forces and torque are measured with a rotary dynamometer in drilling tests for a selected number of holes. Feature engineering is conducted to identify the statistical features of the measurement signals that are most sensitive to small changes in tool wear. A neural network based on the long short-term memory (LSTM) architecture is developed to predict tool flank wear from the measured spindle motor current and dynamometer signals. It is demonstrated that the proposed technique predicts tool flank wear with good accuracy and high computational efficiency. The proposed technique can easily be implemented in an industrial edge device as a real-time predictive maintenance application to minimize the costs due to manufacturing downtime and tool underuse or overuse.

## 1. Introduction

Energy is among the most significant costs in large-scale machining processes in industrial serial manufacturing. Cutting tool wear directly affects energy requirements of machining processes as well as cutting tool costs and the surface quality of the machined part [1]. It is demonstrated that tool wear is positively correlated with power consumption and surface roughness [1, 2]. Furthermore, unplanned tool change in the absence of tool wear monitoring may lead to increased costs due to manufacturing downtime and underuse or overuse of the cutting tool. The prominence of monitoring tool wear in minimizing machining costs has created a demand for real-time techniques applicable to serial manufacturing operations in industry.

Several techniques based on the utilization of multi-sensor data are proposed in the literature for tool wear monitoring. Bagga et al. [3] presented a tool wear prediction technique based on measured vibration, temperature and cutting force data fed into a neural network to predict tool wear during machining. It is demonstrated that the multi-sensor approach predicts tool wear with good accuracy in a range of selected machining conditions. Yan et al. [4] proposed a real-time tool wear monitoring technique utilizing measured acceleration and cutting force signals in conjunction with a convolutional neural network. In addition to the acceleration and cutting force signals, cutting time is also used as an input to the neural network to improve accuracy. It is shown that the technique in [4] predicts tool wear with a maximum error of 8 micrometers. Other tool wear monitoring techniques based on multi-sensor data and the utilization of neural networks are proposed by Gao et al. [5], Xu et al. [6], von Hahn and Mechefske [7] in which various measurement signals such as vibration, cutting force, acoustic emission and spindle motor current are collected as inputs to various neural networks to predict tool wear.

In addition to multi-sensor techniques which rely on measurement signals, several computational techniques are also presented in the literature to simulate tool wear based on physics-based calculations. Urresti et al. [8] proposed a hybrid analytical-numerical technique to simulate tool wear for average thermo-mechanical conditions. A thermal simulation of the cutting operation is carried out based on the finite difference method. The simulated tool wear values are validated against measured tool wear values and it is demonstrated that the simulated results are in good agreement with the experimental results. Other computational techniques in the literature include the semi-analytical modeling technique by Bai et al. [9] based on an abrasive wear model, the thermodynamics-based modeling technique by Bjerke et al. [10] for prediction of wear-related chemical phenomena including oxidation, diffusion and dissolution, and the finite element model by Attanasio et al. [11] based on their tool geometry update subroutine.

Various types of neural networks are used in the literature to predict tool wear. Martínez-Arellano et al. [12] presented a tool wear prediction model based on convolutional neural networks (CNN) and the GASF encoding technique. The authors reported an accuracy of 85% in predicting tool wear using cutting forces, machine vibrations and acoustic emission signals. Wu et al. [13] presented a CNN to monitor the tool wear progression based on camera images of the cutting tool. The authors demonstrated that the proposed method predicted tool wear with 95% accuracy. The recurrent neural network (RNN) is another architecture commonly used in the literature due to its success in predicting sequential events such as the wearing of the cutting tool. Marani et al. [14] used long short-term memory (LSTM) which is an RNN architecture to classify tool health in milling based on spindle motor current signals. The authors demonstrated that the model predicts the progression of tool wear with good accuracy. Wang et al. [15] presented an RNN for tool wear prediction based on the gate-related unit (GRU) architecture to predict tool wear in dry milling based on various time- and frequency-domain features extracted from collected signals. The authors demonstrated that the extracted features effectively compress the initial time-series signals and tool wear is predicted with high accuracy.

Compared to physics-based computational techniques, data-driven multi-sensor tool wear monitoring techniques gained more attention in the literature due to not requiring complex mathematical derivations [16]. Hence, multi-sensor techniques are promising candidates for industrial applications. However, in industrial applications where tolerances are narrow, allowable level of tool wear would be low. This may add difficulties to monitoring efforts due to other components, such as the flexible vibrations of the machine, being more dominant in the measurement signals. In this study, a tool flank wear monitoring technique is introduced to monitor tool flank wear in applications with limited allowable tool wear. The proposed technique uses high-frequency data collected from the machine as inputs to a recurrent neural network based on a long short-term memory (LSTM) architecture for tool flank wear prediction. The proposed technique utilizes feature engineering to identify the statistical characteristics of the measurement signals that are most sensitive to tool wear. The selected statistical features are further processed to highlight trends in the measurement signals which improves prediction accuracy while significantly reducing model training time. The proposed tool wear monitoring technique is easy to implement in industrial applications with limited computational resources and predicts tool wear with good accuracy.

## 2. Drilling Experiments

Drilling experiments are conducted to generate the training and validation data sets for the proposed tool flank wear prediction algorithm. The test setup is shown in Fig. 1. The drilling tests are conducted on a Spinner U1530 five-axis machine tool with a Siemens SINUMERIK 840D sl controller. SINUMERIK Edge, an industrial edge device by Siemens for Internet of Things (IoT) applications, is used to collect the spindle motor current with a sampling frequency of 500 Hz [17]. The cutting torque and forces are collected with a Kistler 9123CQ05 rotary dynamometer attached to the spindle nose. The dynamometer measurement signals are first amplified with a Kistler Type 5223 charge amplifier and digitized with an NI USB-6259 data acquisition system by National Instruments. The data acquisition hardware is controlled by the CutPro software by Manufacturing Automation Laboratories Inc.

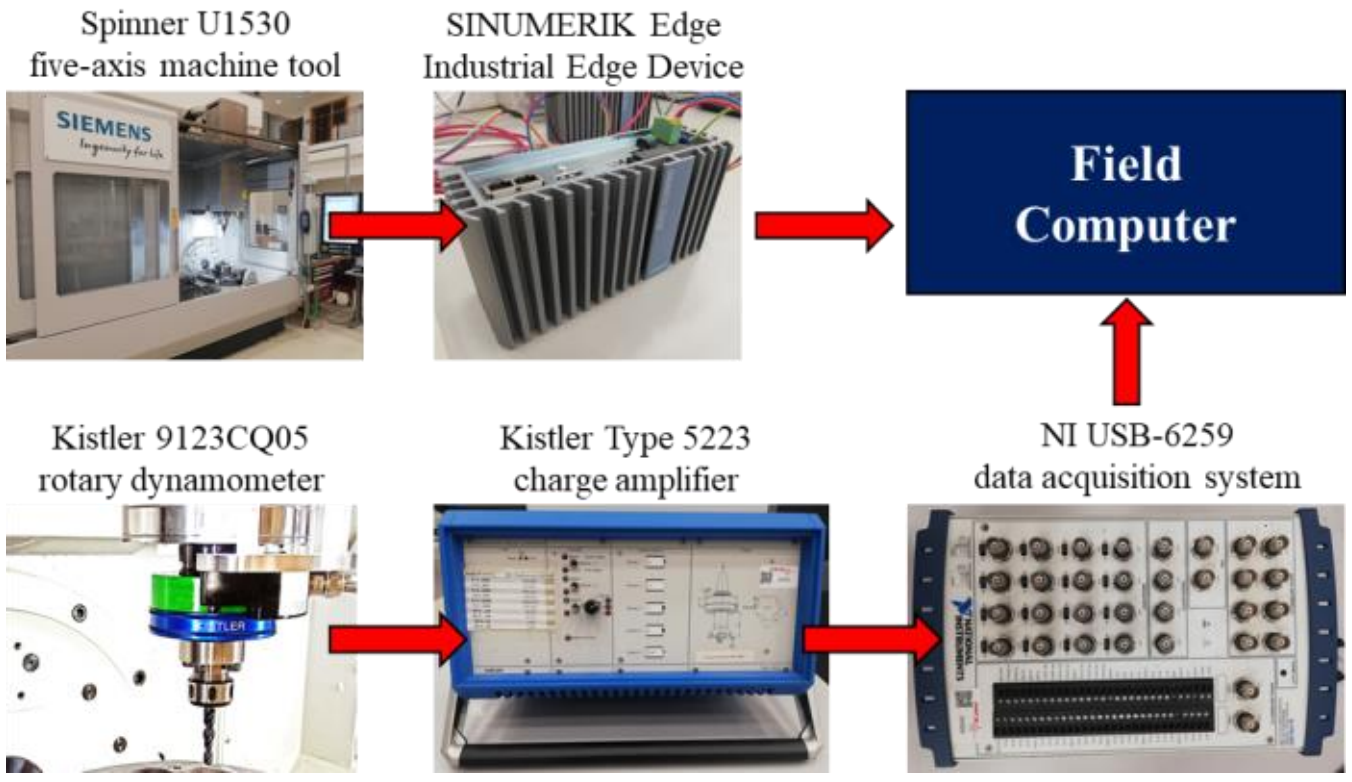

Fig. 1. The test setup used for tool flank wear experiments.

AISI 1045 carbon steel blocks with dimensions 400x325x25 mm are used as the workpiece. A TiAlN-coated carbide twist drill with two flutes, 8.5 mm diameter and 140° tip angle is used as the tool. A total of 1901 holes are drilled on six workpieces as shown in Fig. 2. The spindle speed and the feed rate are selected as 2400 revolutions per minute (rpm) and 400 mm/min, respectively.

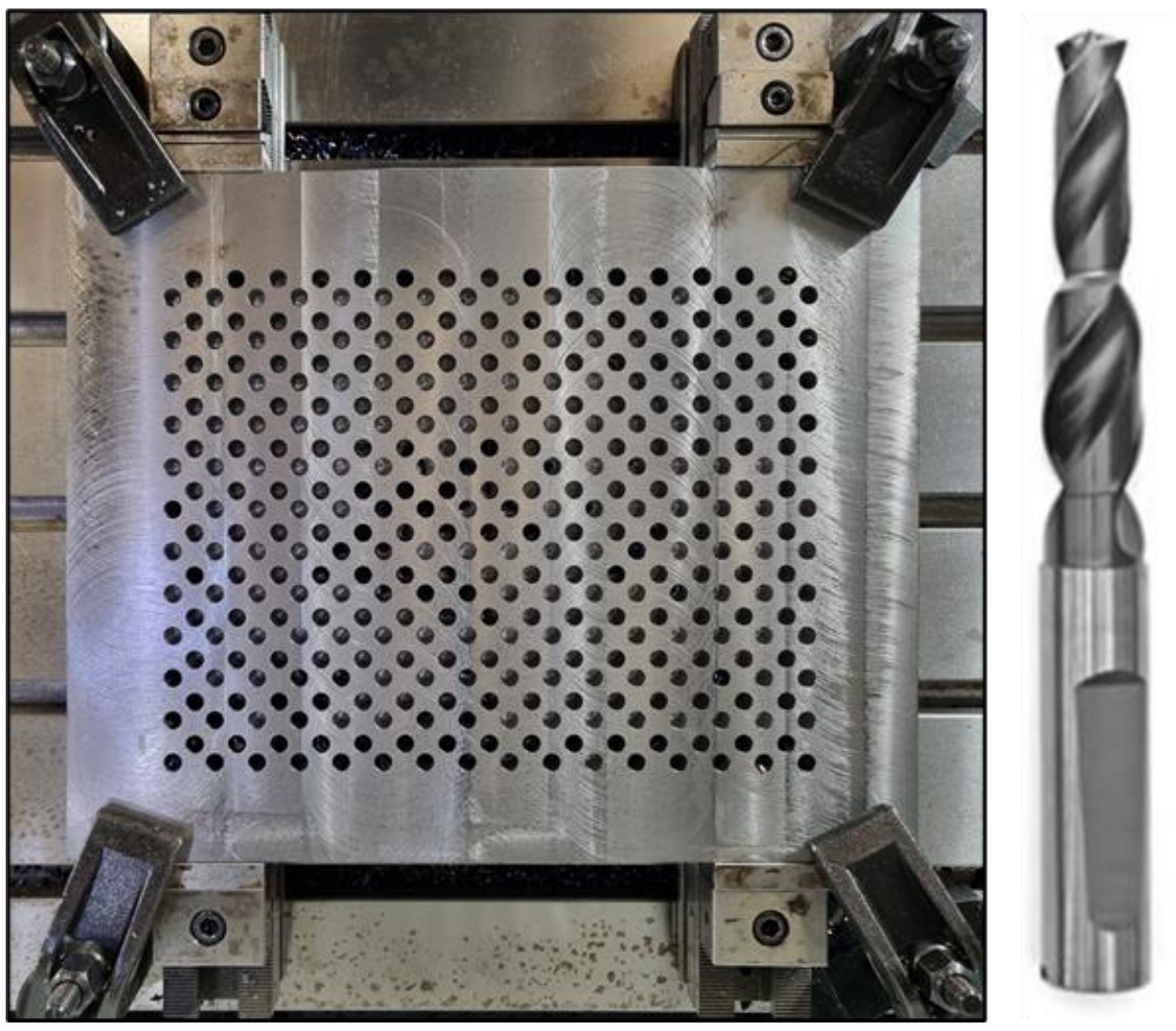

Fig. 2. The workpiece, drilling tool and the hole pattern used.

While the spindle motor current, the cutting torque and the cutting forces are collected simultaneously for each hole with the edge device and the dynamometer, tool flank wear is measured once for every 48 holes to keep the total test time within feasible limits. After each group of 48 holes is drilled, the machine is stopped and the tool is detached from the dynamometer to be taken to the microscope for tool flank wear measurements. The microscope images of both flutes are captured with the software and the tool is attached to the dynamometer to drill the next set of 48 holes while the resulting wear is measured in micrometers based on the microscope images. After all the steel blocks are drilled, a total of 1901 samples are obtained for the spindle motor current, the cutting torque and the cutting force corresponding to each hole while 40 tool flank wear values are obtained depicting the evolution of tool flank wear through the process in Fig. 3.

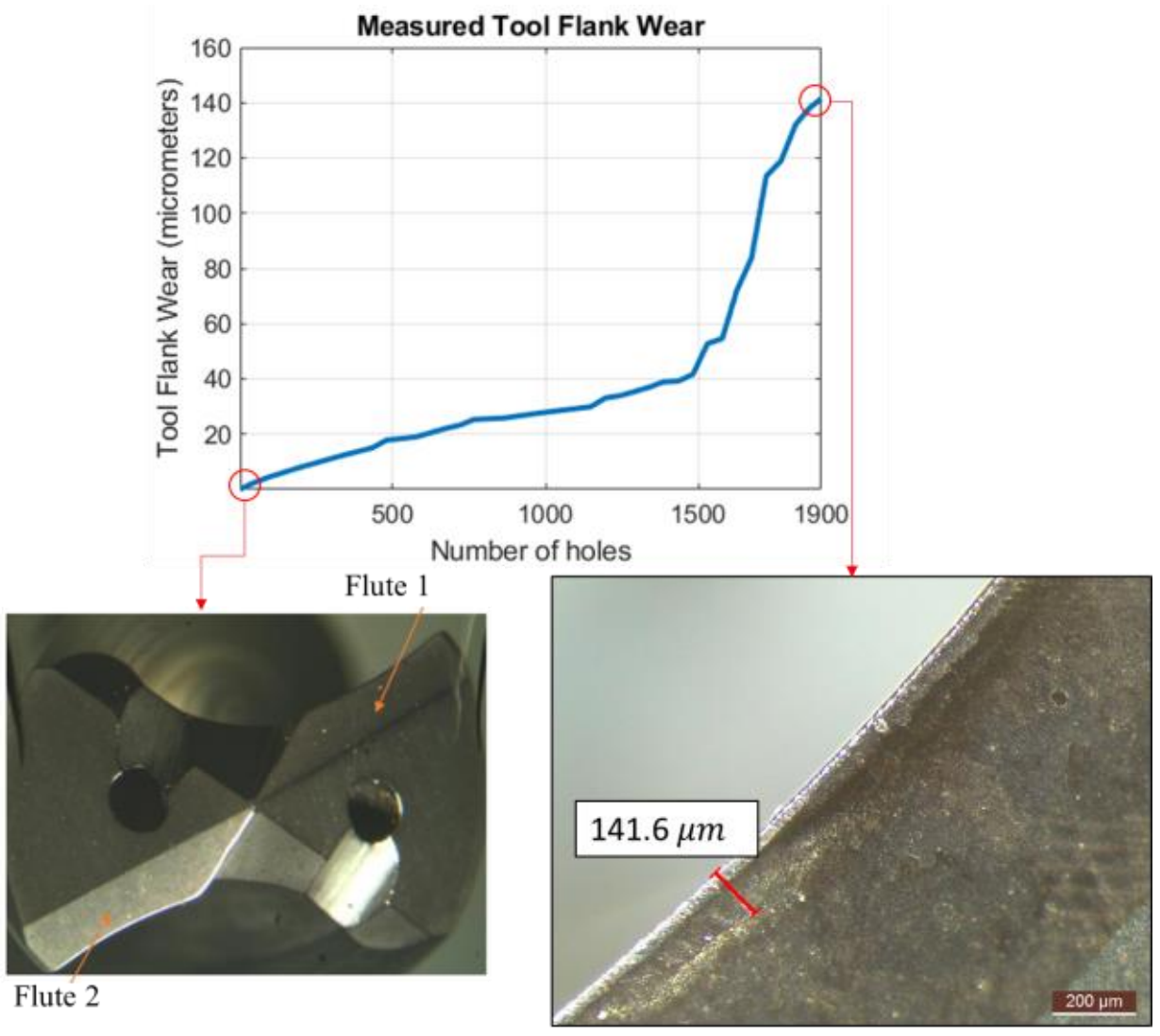

Fig. 3. Tool flank wear curve obtained from the drilling experiments.

## 3. Tool Flank Wear Prediction using Deep Learning

The model development steps of the proposed tool wear prediction method consist of below procedures:

1. Pre-processing of the raw measurement data for order reduction and cleanup
2. Identification and processing of the statistical features for improved accuracy and efficiency
3. Recurrent neural network (RNN) training and testing for tool wear prediction

Below sections explain the details of these steps contributing to a good tool wear prediction accuracy.

### *3.1. Pre-processing of raw experimental data*

The raw measurement signals obtained from the industrial edge device and the dynamometer contain information obtained both in and out of the cutting events. The signals collected during the drilling of each hole (i.e., the cutting signals) contain the most essential information for further analysis while the signals collected as the tool moves from one hole to the next are not useful. Additionally, the cutting signals contain undesirable components due to factors such as electrical noise, friction, and transient vibrations. Hence, after the cutting signals are isolated from the raw measurement data, they are processed for improved prediction accuracy. Low-pass or band-pass filtering in the frequency domain are widely adopted techniques in the literature to remove undesirable components from measurement signals. Although these techniques are often successful in clearing raw measurement signals, the resulting signals have the same number of data points as the original ones. The number of data points can be on the order of hundreds or thousands depending on the selected sampling frequency and the duration of the machining process.

Using filtered signals comprising hundreds or thousands of data points as inputs to deep learning algorithms may significantly increase the computational resources and time required for model training and testing which may exceed the available resources in an industrial environment. Furthermore, the resulting model may fail to achieve the capability to produce real-time tool wear predictions as a result of the increased computational demand. In this study, a set of statistical properties of the collected time-domain cutting signals which comprise only a few scalars are used as inputs to an RNN instead of the hundreds of data points constituting the cutting signals. The scalars are selected such that the cutting signals' statistical properties in the time domain are reflected as features in the recurrent neural networks for improved tool wear prediction accuracy. The initially selected statistical features are the root mean square (RMS), the standard deviation (STD), and the spectral power (SPW) of the cutting signals. The initial set of statistical features are calculated using equations (1), (2), and (3).

$$y_{RMS} = \sqrt{\frac{1}{N}\sum_{i=1}^{N} y_i^2} \quad (1)$$

$$y_{STD} = \sqrt{\frac{1}{N}\sum_{i=1}^{N} (y_i - y_{mean})^2} \quad (2)$$

$$y_{SPW} = \frac{1}{2}\sum_{i=\omega_{start}}^{\omega_{end}} |Y(i)|^2 \quad (3)$$

As tool wear is measured for only 40 of the 1901 holes, the 40 tool wear measurement points are quantized to a total of 1901 tool wear values. Initially, an appropriate number of tool wear values are calculated based on linear interpolation between two subsequent tool wear measurements. Then, a random variation is added to the linearly interpolated tool wear values to mimic the complex tool wear progression in reality as well as the uncertainties due to the measurement device. For example, if two subsequent tool wear measurements are made for holes number 1824 and 1872, 46 new values are calculated by linearly interpolating the values of tool wear measured for holes number 1824 and 1872. Then, a random number generator is used to add random variations to each of these 46 tool wear quantizations. The obtained quantizations correspond to the tool wear approximations for holes number 1825 – 1871. Some tool wear values obtained from actual measurements and quantization are shown in Fig. 4.

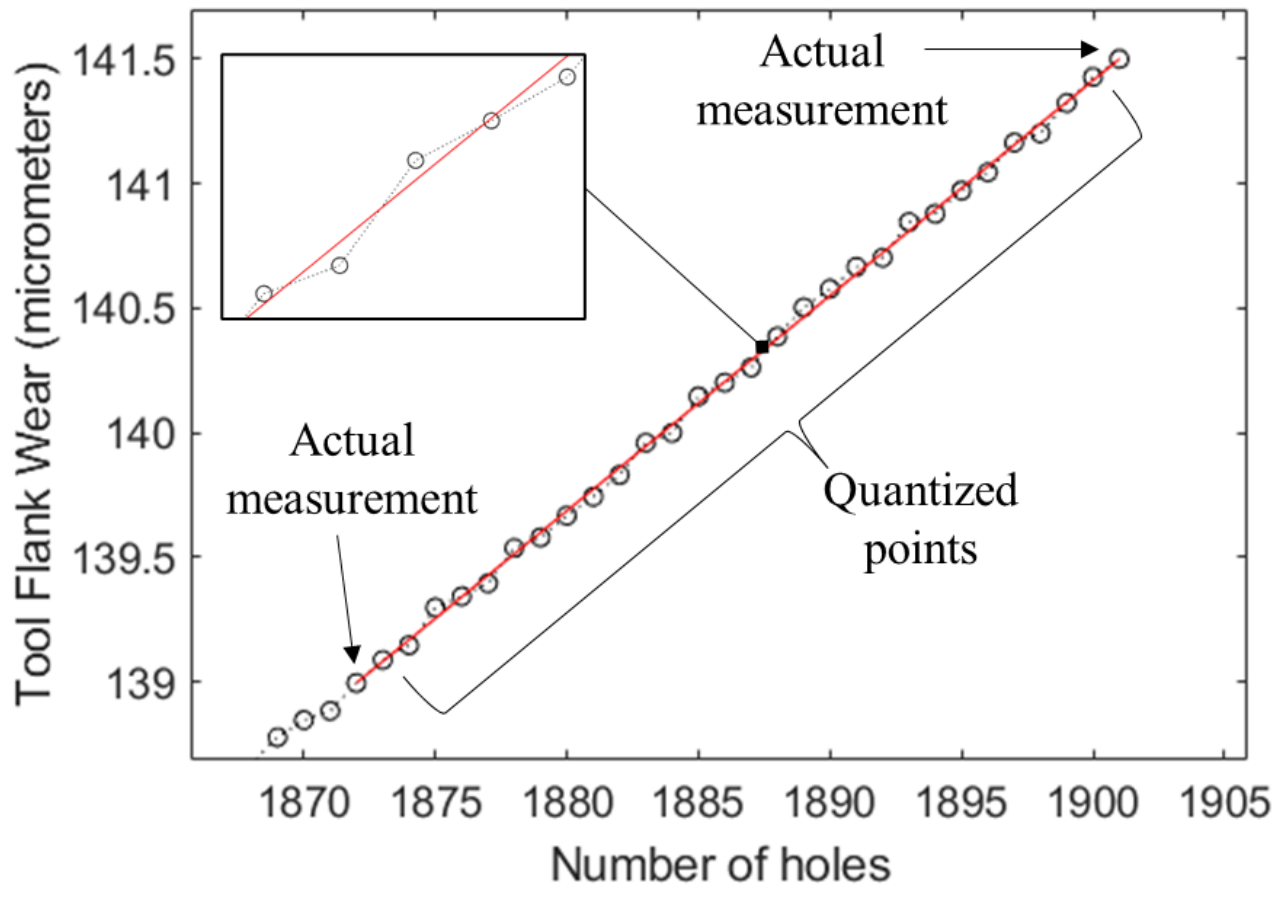

Fig. 4. Quantized tool flank wear curve segment.

## 3.2. Identifying and processing statistical features

The initially selected statistical features given by equations (1), (2) and (3) are calculated individually for the spindle motor current (Im), the cutting force in Z axis (Fz) and the cutting torque (Tz) as shown by the blue lines in Fig. 5. Although the flank wear curve in Fig. 3 has a constant upward trend, Fig. 5 shows that the statistical features have varying trends which is due to the complex wear mechanism of the coated tool. To identify the changing trends in the statistical features, the moving average, as shown by the red lines in Fig. 5, is calculated for each feature with a window size of 200 samples.

To maximize the prediction accuracy despite the changing profile of the moving average of the features, it is proposed to split the measurement data into regions where the moving average has less variance in its trend. One way the data set can be split into regions to separate varying trends is shown in Fig. 5 where the first, the second, and the third regions contain data from holes 200 – 800, 800 – 1400, and 1400 – 1800, respectively. The data segment after the 1800th hole is discarded as the window size of the moving average gets considerably smaller towards the end of the data set, distorting the averaged signal. These regions are named Region 1, Region 2, and Region 3, respectively. As seen by the variation of the moving averages in Fig. 5, while the RMS and the STD are responsive to advancing tool wear especially in Region 1, the spectral power (SPW) mostly follows a flat profile, indicating a low sensitivity for increasing tool wear. Hence, SPW is discarded from the feature set and only RMS and STD are used to train and test the tool flank wear prediction models.

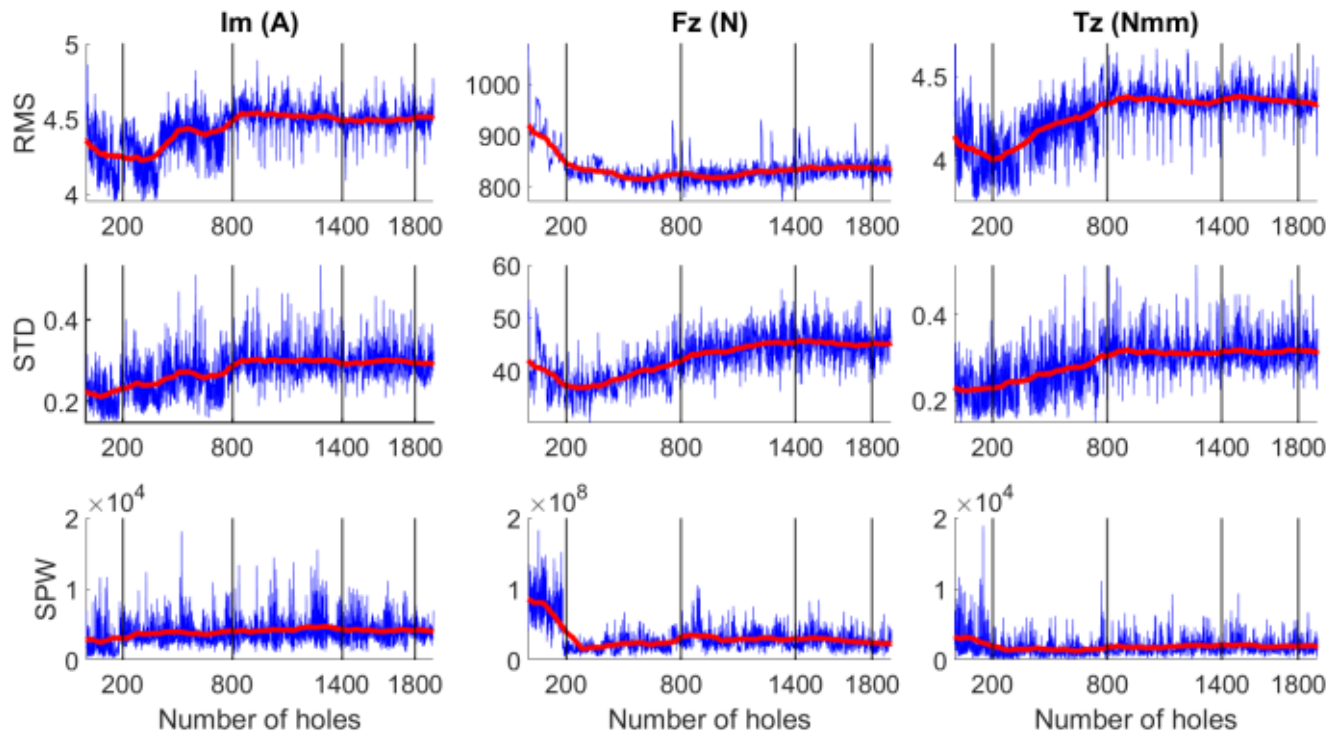

Fig. 5. Statistical features (blue) and their moving averages (red).

## 3.3. Model training and tool wear prediction with LSTM

Tool wear is an incremental event in the physical world. In the case of drilling, the resulting tool wear after a new hole is drilled is built upon the existing tool wear prior to the drilling of the new hole. LSTM (long short-term memory) is a recurrent neural network architecture that is successful in predicting sequential events [18]. Hence, LSTM is selected to predict the incremental nature of tool wear.

As seen in Fig. 6, The full sample set of 1901 holes is divided into training, validation, and test sets by using the ratios 75%, 15%, 10%, respectively. The LSTM timestep is selected as 20 to allow the LSTM algorithm to use the information from the past 20 holes to predict the tool wear at the end of the 20th hole. The total number of features for each sample, corresponding to each hole, is 6 as there are three selected signals (i.e., the spindle motor current, the cutting force in Z direction, the cutting torque) and two selected statistical features (i.e., RMS and STD) for each signal. The target vector is selected as the quantized tool flank wear vector partially graphed in Fig. 4. To automatically search for an optimal set of hyperparameters for each of the Region 1, Region 2 and Region 2 neural networks, Keras Tuner [19] is used with a range of 1 – 10 layers and 16 – 128 neurons per layer, rectified linear unit (ReLU) or hyperbolic tangent (tanh) as activation functions, a range of dropout and recurrent dropout rates between 0.0 and 0.5, L1 or L2 kernel regularizers with various factors, and a range of learning rates between 1E-4 and 1E-2 for the Adam optimizer. Mean absolute error is used as the loss. The tuner is run for 100 epochs with an early stopping option if the value of loss function remains similar for 10 consecutive epochs.

As seen in Fig. 5, the scales of the features differ substantially which is found to harm the learning process. Hence, after the sample set is divided into training, validation and test sets, all three sets are standardized with respect to the training set by using a min-max scaler to scale the training sets to the range 0.0 – 1.0. The mean absolute percentage error (MAPE) is selected as the performance criterion for the tool wear predictions with the test set. MAPE calculates the average error between the measured and predicted sets and reflects the neural network's accuracy in physical terms.

## 4. Results and Discussion

The training, validation, and prediction results are shown in comparison with the measured tool flank wear curve in Fig. 6. As seen, good prediction accuracy is achieved in all three regions covering distant levels of wear along the quantized tool wear curve with a maximum MAPE of 6.6%. Additionally, using the proposed technique, the Keras Tuner with the Hyperband algorithm trains 300 models in an hour on a computer with an AMD Ryzen 5 2600 6-Core CPU and 32 GB RAM, which corresponds to about 10 seconds per model. It is seen that when the full time-domain data comprising hundreds of data points is used as input instead of the proposed moving averaged statistical features, the same computer trains one model in 15 minutes. Therefore, the proposed technique can predict tool flank wear with good accuracy while being computationally feasible for industrial projects with limited computational resources.

Although this paper presents the proposed model as a regression model to predict the value of tool wear given a set of edge device and dynamometer signals, it is inferred from Fig. 6 that the proposed architecture can also be used for classification to predict the region representing the tool's state which may provide sufficient information in some predictive maintenance applications. For both the regression and classification implementations of the proposed technique, the number of regions may be increased at a low additional computational cost to improve accuracy and resolution.

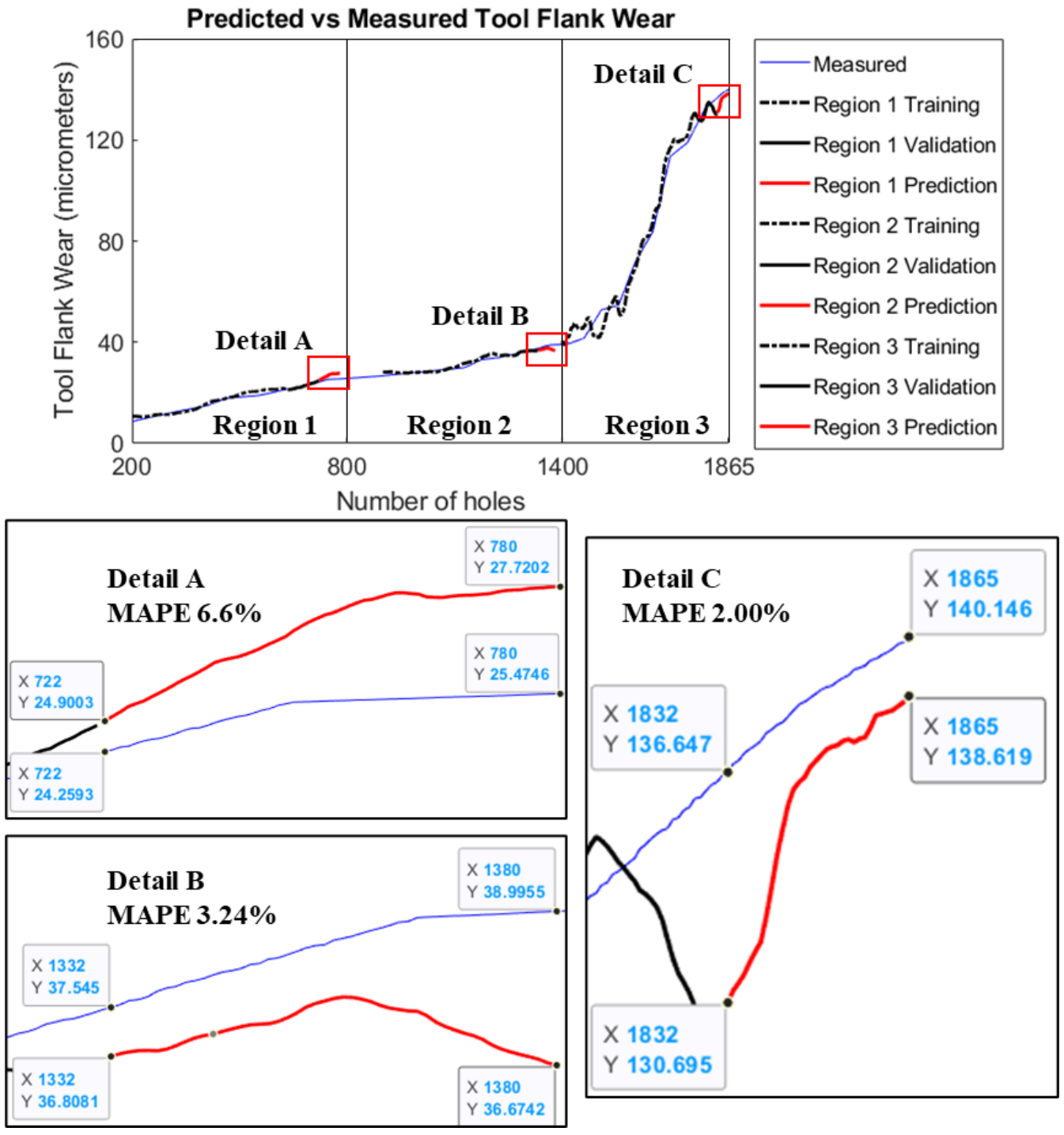

Fig. 6. Comparison of measured (quantized) and predicted tool flank wear.

## 5. Conclusion

In this study, an accurate and computationally efficient tool flank wear prediction technique is presented based on a recurrent neural network using spindle motor current and dynamometer signals as inputs. Instead of the full time-domain vectors of the measurement signals comprising hundreds of data points, their representative averaged statistical features consisting of a few scalars are used as inputs for prediction, which significantly reduced training times and improved training performance. A series of drilling tests are performed and the cutting signals are simultaneously collected with the combined use of an industrial edge device and a rotary dynamometer. The features are inspected for sensitivity to the level of tool flank wear based on a moving average approach and the least sensitive features are discarded from the input data set. Three recurrent neural networks based on the LSTM (long short-term memory) architecture are created to predict tool flank wear at distant regions on the tool wear curve. It is demonstrated that the proposed LSTM model predicts tool flank wear with good accuracy while improving computational efficiency. The proposed tool flank wear prediction technique is easy to implement in industrial settings with limited time and computational resources.

## Acknowledgements

The authors would like to thank Ford Otosan and Siemens for their support in obtaining necessary hardware and materials. The authors would also like to thank Siemens IoT Edge Research Lab in the Manufacturing and Automation Research Center (MARC) at Koç University for their support with all necessary testing and measurements.

## Author Statement

The authors would like to state that the first four authors have contributed equally to this research.